\documentclass[a4paper,12pt]{article}
\usepackage{jheppub_modified_green}

\usepackage{epsfig}
\usepackage{amsfonts}
\usepackage{amssymb}
\usepackage{enumitem}
\usepackage{amsthm}
\usepackage{subfig}
\usepackage{bm}
\usepackage{hyperref}
\usepackage{mathrsfs}
\usepackage{bbm}
\usepackage{cancel}
\usepackage{xcolor}
\usepackage{accents}
\usepackage{relsize}
\usepackage{float, slashed, graphicx, amssymb, amsmath}
\usepackage{shuffle}
\usepackage{tikz}
\usetikzlibrary{hobby,calc,shapes, positioning, backgrounds,trees, decorations, decorations.markings, decorations.pathmorphing, arrows, shapes.geometric, snakes}
\usepackage{tikz-feynman}
\tikzfeynmanset{compat=1.1.0}
\tikzfeynmanset{
graviton/.style={circle, draw=green!60, fill=green!5, very thick, minimum size=7mm}
}
\tikzfeynmanset{
codot/.style={/tikz/shape=circle,
/tikz/fill=red,/tikz/minimum size=0.1cm,/tikz/inner sep=1.8pt}
}
\tikzfeynmanset{sb/.style=
{/tikz/shape=circle,
/tikz/fill=black,
 /tikz/minimum size=0.3cm,/tikz/inner sep=1.pt
 } }
\tikzfeynmanset{myblob/.style=
{/tikz/shape=ellipse,
/tikz/fill=red,
 /tikz/minimum width=0.5cm,
 } }
 \tikzfeynmanset{myblobFF/.style=
{/tikz/shape=circle,
/tikz/fill=black,
 /tikz/minimum width=0.25cm,
 } }
 \tikzfeynmanset{myblob2/.style=
{/tikz/shape=rectangle,
/tikz/fill=black,
 /tikz/minimum width=0.1cm,/tikz/inner sep=1.8pt
 } }
  \tikzfeynmanset{fp/.style=
{/tikz/shape=rectangle,
/tikz/fill=red,
 /tikz/minimum width=0.1cm,/tikz/inner sep=1.8pt
 } }
  \tikzfeynmanset{myvertex/.style=
{/tikz/shape=circle,
/tikz/fill=black,
 /tikz/minimum width=0.05cm,
 } }
\tikzset{box/.pic={\filldraw[fill=black]  (0,0) circle (2.5pt);
				   \filldraw [fill=black] (0.5,0) circle (2.5pt);
			       \draw [line width=5pt] (0,0) -- (0.5,0);}}

\tikzset{wiggle/.style={decorate, decoration=snake}}

 \tikzfeynmanset{HV/.style=
{/tikz/shape=circle,
/tikz/fill={rgb:black,1;white,2},
 /tikz/minimum size=0.01cm,/tikz/inner sep=1.8pt
 } }

\usepackage[T1]{fontenc} 

\newcommand{\tr}{\text{tr}}

\def\veps{\varepsilon}
\def\sc#1{\overline{#1}}
\makeatletter
\newcommand \UPlus {\mathop {\operator@font \uplus }\limits }
\makeatother
\makeatletter
\newcommand \Bigcup {\mathop {\operator@font \bigcup }\limits }
\makeatother
  \def\LabelNote#1{}
 \def\Label#1{\label{#1}%
  \smash{\hbox to\phipt{\raise1ex\hbox{\tiny[#1]}\hss}}}

\def\nn{\nonumber}

\newcommand{\red}{\color{red}}

\def\spa#1.#2{\left\langle#1\,#2\right\rangle}
\def\spb#1.#2{\left[#1\,#2\right]}

\def\be{\begin{equation}}
\def\ee{\end{equation}}
\def\bea{\begin{eqnarray}}
\def\eea{\end{eqnarray}}  
\allowdisplaybreaks

\newcommand{\neco}{\mathbb{L}}          
\newcommand{\npre}{\mathcal{N}}

\newcommand{\la}{\langle}
\newcommand{\ra}{\rangle}
\newcommand{\mdot}{{\cdot}}

\usepackage{hyperref}
\hypersetup{
	colorlinks=true,
	linktoc=page,
	citecolor=americanrose,
	linkcolor=cadmiumgreen,
	urlcolor=blue} 
\definecolor{americanrose}{rgb}{1.0, 0.01, 0.24}
\definecolor{cadmiumgreen}{rgb}{0.0, 0.42, 0.24}

\title{Kinematic Hopf algebra for amplitudes from higher-derivative operators}
\author{Gang Chen$\mbox{}^{a}$,}
\author{Laurentiu Rodina$\mbox{}^{b,c}$,}
\author{Congkao Wen$\mbox{}^{b}$\\}
\affiliation{$\mbox{}^{a}$Niels Bohr International Academy,
Niels Bohr Institute, University of Copenhagen,\\
Blegdamsvej 17, DK-2100 Copenhagen \O, Denmark}
\affiliation{$\mbox{}^{b}$Centre for Theoretical Physics, Department of Physics and Astronomy, \\
Queen Mary University of London, Mile End Road, London E1 4NS, United Kingdom}
\affiliation{$\mbox{}^{c}$Beijing Institute of Mathematical Sciences and Applications (BIMSA), Beijing, 101408, China}

\emailAdd{gang.chen@nbi.ku.dk}
\emailAdd{laurentiu.rodina@gmail.com}
\emailAdd{c.wen@qmul.ac.uk}
\begin{document}
\begin{flushright}
	QMUL-PH-23-22
\end{flushright}

\abstract{
Recently it has been shown that Bern-Carrasco-Johansson (BCJ) numerators of colour-kinematic duality for tree-level scattering amplitudes in Yang-Mills theory (coupled with scalars)  can be determined using a quasi-shuffle Hopf algebra. In this paper we consider the same theory, but with higher-derivative corrections of the forms $\alpha' F^3$ and $\alpha'^2 F^4$, where $F$ is the field strength. In the heavy mass limit of the scalars, we show that the BCJ numerators of these higher-derivative theories are governed by the same Hopf algebra. In particular, the kinematic algebraic structure is unaltered and the derivative corrections only arise when mapping the abstract algebraic generators to physical BCJ numerators. The underlying kinematic Hopf algebra enables us to obtain a compact expression for the BCJ numerators of any number of gluons and two heavy scalars for amplitudes with higher-derivative operators. The pure gluon BCJ numerators can also be obtained from our results by a simple factorisation limit where the massive particles decouple. 
}


\maketitle

\tableofcontents

\section{Introduction}

The Bern-Carrasco-Johansson (BCJ) duality reveals deep connections between  Yang-Mills theory and gravity \cite{Bern:2008qj,Bern:2010ue}, and provides remarkably powerful computational tools for calculating scattering amplitudes including at loop level. See \cite{Bern:2019prr,Bern:2022wqg} for recent reviews and the references therein. 
A central aspect of the duality is the fact that BCJ kinematic numerators of scattering amplitudes satisfy non-trivial relations, in particular the Jacobi identity associated with the dual colour factors. This extremely intriguing fact has not been understood in full generality. Given both the practical and theoretical implications of the duality, understanding the kinematic algebra that is responsible for the duality remains an important open question.

As an abstract algebra, it is generally characterised in terms of fusion products for a basis of generators. The generators can be realised from several different points of view, such as differential operators \cite{Monteiro:2011pc,Ben-Shahar:2021zww,Cheung:2021zvb,Monteiro:2022nqt,Monteiro:2022lwm,Lipstein:2023pih}, auxiliary fields \cite{Cheung:2016prv,Ferrero:2020vww,Ben-Shahar:2022ixa}, string worldsheet operators \cite{Mafra:2022wml,Fu:2018hpu,Ben-Shahar:2021doh}, or BV-BRST formalism \cite{Borsten:2020zgj,Borsten:2020xbt,Borsten:2022vtg, Bonezzi:2023pox, Borsten:2023reb, Borsten:2023ned, Borsten:2023paw}. Recently a systematic framework for the kinematic algebra was proposed in \cite{Chen:2019ywi,Chen:2021chy,Brandhuber:2021kpo}, where the generator is taken as the QCD current, or its heavy-mass limit.  Especially in the heavy-mass effective field theory \cite{Georgi:1990um, Luke:1992cs, Neubert:1993mb, Manohar:2000dt,Damgaard:2019lfh,Brandhuber:2021kpo,Brandhuber:2021eyq}, which had a wide phenomenological applications in heavy quark  and black hole physics \cite{Brandhuber:2021eyq,Brandhuber:2023hhy,Bjerrum-Bohr:2023iey,Bjerrum-Bohr:2023jau,Brandhuber:2023hhl,Herderschee:2023fxh,Haddad:2023ylx,Aoude:2020onz,Aoude:2023vdk,Adamo:2022ooq}, the kinematic algebra was found to be isomorphic to  an infinite dimensional combinatorial algebra \cite{Brandhuber:2021bsf,Brandhuber:2022enp}, i.e. a (generalised) quasi-shuffle Hopf algebra \cite{hoffman2000quasi,Blumlein:2003gb,aguiar2010monoidal,hoffman2017quasi,fauvet2017hopf}. This underlying kinematic algebra, which is called the kinematic Hopf algebra, was further extended to scalar Yang-Mills theory  with an arbitrary finite mass in \cite{Chen:2022nei}. A geometrical perspective of this structure was also proposed in \cite{Cao:2022vou}.

A natural question is to find out whether all theories that admit a colour-kinematic representation also admit such a Hopf algebra structure. In this paper we target higher derivative corrections to Yang-Mills theory and find the answer is positive. Specifically, we consider the subset of higher-derivative operators of the forms $\alpha' F^3$ and $\alpha'^2 F^4$  that are compatible with adjoint type colour-kinematic duality, as discussed in \cite{Broedel:2012rc}. Such higher-derivative operators are relevant for the bosonic string amplitude in the low energy expansion.

We find the algebra structure (the fusion product rules, etc.) is in fact identical to the previously studied cases, and the only modification required is to the map which turns the algebraic generators into concrete functions of momenta and polarisation vectors -- such map is called the evaluation map \cite{Chen:2022nei}. We provide explicit  expressions for the $\alpha'$ corrections to the evaluation maps due to $\alpha' F^3, \alpha'^2 F^4$ terms. The underlying Hopf algebra structure enables us to write down extremely compact expressions for the BCJ numerators of the higher-derivative theory for the amplitudes with any number of gluons and two heavy scalars. The BCJ numerators of pure gluons can be also obtained from our results by an appropriate factorisation limit in which the scalars decouple \cite{Brandhuber:2021bsf}. We have explicitly verified the BCJ numerators obtained from the proposed kinematic algebra  up to eight external particles.

The rest of the paper is organised as follows. In Section \ref{HD} we introduce the relevant higher derivative corrections to Yang-Mills with $\alpha' F^3$ and $\alpha'^2 F^4$ terms, and describe the procedure resulting in the corresponding BCJ numerators with two heavy scalars. In Section \ref{KH2} we briefly review the kinematic Hopf algebra construction of BCJ numerators by introducing fusion product rules of Hopf algebra and the concept of evaluation maps. In Section \ref{KH} we propose an explicit expression for the evaluation maps that compute BCJ numerators of all multiplicity up to the order $\alpha'^2$.  We conclude in Section \ref{conc} with a summary and possible future directions.  Finally, Appendix \ref{sec:sixpts} contains the explicit expression for the numerators with six particles, and higher-point expressions can be found in the linked repository~\cite{ChenGitHub}.

\section{Higher-derivative corrections  and BCJ numerators}\label{HD}

In \cite{Broedel:2012rc} it was pointed out that up to order $\alpha'^2$ there exists a unique combination of higher derivative operators in Yang-Mills theory that is compatible with adjoint type colour-kinematic duality. In fact, it was recently understood that the entire series of higher derivative corrections is tightly constrained by demanding ``double copy consistency" \cite{Carrasco:2022lbm}. The Lagrangian that is consistent with  the adjoint type colour-kinematic duality up to order $\alpha'^2$ takes the following form, 
\begin{equation} \label{eq:Lag}
S_{\mathrm{YM}+\alpha'F^3+\alpha'^2F^4}=\int \mathrm{d}^D x \operatorname{Tr}\left\{\frac{1}{4} F_{\mu \nu} F^{\mu \nu}+\frac{2 \alpha^{\prime}}{3} F_\mu^\nu F_\nu^\lambda F_\lambda^\mu+\frac{\alpha^{\prime 2}}{4}\left[F_{\mu \nu}, F_{\lambda \rho}\right]\left[F^{\mu \nu}, F^{\lambda \rho}\right]\right\}\,,
\end{equation}
and this is the theory we will study in this paper. 

An efficient method to compute the corresponding amplitudes based on Berends-Giele recursion \cite{Berends:1987me} was proposed in~\cite{Garozzo:2018uzj}. We will be interested in scattering amplitudes of gluons coupled to two heavy scalars, which have been shown to play very important roles in the study of gravitational wave emission and black hole physics \cite{Brandhuber:2021eyq,Brandhuber:2023hhy,Bjerrum-Bohr:2023iey,Bjerrum-Bohr:2023jau,Brandhuber:2023hhl,Herderschee:2023fxh,Haddad:2023ylx,Aoude:2020onz,Aoude:2023vdk,Adamo:2022ooq}. One way to obtain these gluon-scalar amplitudes is by performing a dimensional reduction of two gluons to two massive scalars  at the level of Lagrangian, starting from \eqref{eq:Lag}. Effectively, one can directly perform the dimensional reduction at the level of scattering amplitudes using the transmutation operator $\partial_{e_i}\mdot \partial_{e_j}$ following~\cite{Cheung:2017ems} (see also \cite{Chiodaroli:2015rdg}), which turns two gluons into adjoint scalars. We then take the infinite mass limit for these scalars, which have the momentum $m v^\mu$, with $m$ and $v^\mu$ being the mass and  velocity of the heavy particle.

The corresponding BCJ numerators of these amplitudes can be obtained through the KLT procedure \cite{Kawai:1985xq}: 
\begin{equation} \label{eq:KLT}
\mathcal{N}_{\mathrm{BCJ}}(1 \alpha, v)=\sum_{\beta \in S_{n-3}} \mathcal{S}(1 \alpha \mid 1 \beta) A(1 \beta, v)\,,
\end{equation}
where $A(1 \beta, v)$ is the amplitude of $n-2$ gluons and two heavy scalars which can be obtained from the procedure outlined above, and $\mathcal{S}(1 \alpha \mid 1 \beta)$ is the KLT matrix \cite{BjerrumBohr:2010hn}. For example, using \eqref{eq:KLT}, and taking the heavy mass limit, we find the  numerators for the three and four-point amplitudes determined by the Lagrangian \eqref{eq:Lag} can be expressed as
\begin{align} 
\mathcal{N}_{\mathrm{BCJ}}(1, v)= \varepsilon_1\mdot v\, , 
\end{align}
which cannot receive any high-derivative corrections when two of external particles are scalars, and
\begin{align} \label{eq:4pts}
 \mathcal{N}_{\mathrm{BCJ}}(12,v)=&-\frac{v\mdot F_{12} \mdot v}{p_1\mdot v}+\alpha' \tr(F_{12}) p_1\mdot v+\alpha'^2 p_{12}^2 \tr(F_{12}) p_1\mdot v\,,
\end{align}
where $F^{\mu\nu}_i=p_i^\mu \veps_i^\nu -\veps_i^\mu p_i^\nu$, and we have defined, 
\begin{align}
F_{\sigma} = F_{i_1} \cdot F_{i_2} \cdots  F_{i_r}\, ,\, p_{\sigma}=p_{i_1}+p_{i_2}+\cdots +p_{i_r}, \quad {\rm for} \quad \{i_1, i_2, \ldots, i_r\} =\sigma \, ,
\end{align}
 in particular $F_{12} = F_1\mdot F_2, p_{12}=p_1+p_2$. The higher derivative corrections are particularly simple here, but naively they grow quickly in complexity at higher multiplicity.  In the next section, we will present a Hopf algebra that underlies the mathematical structures of these BCJ numerators with higher derivative corrections. The understanding of the algebraic structures allows us to write compact expressions for the numerators with arbitrary multiplicity.

It is worth mentioning that once the numerators are obtained, applying the double copy procedure leads to potential higher derivative corrections in gravity \cite{Broedel:2012rc, He:2016iqi, Garozzo:2018uzj}. In particular, by squaring the four-point numerator given in \eqref{eq:4pts}, we find the contribution arising from $F^3 \times F^3$ is given by
\begin{align}
  \alpha'^2  \frac{( \tr (F_1 \mdot F_2) \, p_1 \mdot v)^2}{p^2_{12}} \, , 
\end{align}
which  agrees with the result given \cite{Brandhuber:2019qpg}  for the gravitational amplitude of two gravitons and two heavy scalars with one insertion of $R^3$. The double copy of $F^3 \times F^3$ should also include the contribution from the insertions of two $R^2 \phi$, which however is proportional to that of $R^3$ at four points, which explains the agreement with the amplitude of a single $R^3$ insertion.  

\section{Review of Kinematic Hopf algebra}\label{KH2}

In this section we will give a brief review of constructing BCJ numerators based on the Kinematic Hopf algebra  \cite{Brandhuber:2021bsf,Chen:2022nei, Brandhuber:2022enp}. The construction consists of two steps. The first step is to introduce the quasi-shuffle Hopf algebra of algebraic generators, whereas the second step is to determine the evaluation map that translates algebraic generators into functions of kinematics (i.e. momenta and polarization vectors). 

\subsection{Review of quasi-shuffle Hopf algebra}

We begin by introducing the quai-shuffle Hopf algebra relevant for constructing BCJ numerators, following the references \cite{Brandhuber:2021bsf,Chen:2022nei, Brandhuber:2022enp}. More details can also be found in \cite{hoffman2000quasi,Blumlein:2003gb,aguiar2010monoidal,hoffman2017quasi,fauvet2017hopf}. We first construct an abstract algebra numerator which is obtained directly from the fusion product of the algebraic  generators  $T_{(i)}$ as follows, 
\begin{align} \label{eq:Nhat}
    \widehat\npre(12\ldots n{-}2, v)=T_{(1)}\star T_{(2)}\cdots \star T_{(n{-}2)}\, ,
\end{align}
with the fusion product defined as the standard quasi-shuffle product. 
For instance, for the cases with two, three, and four gluons, we have 
\begin{align}
    \widehat\npre(12, v)=&\,\, T_{(1)}\star T_{(2)}=-T_{(12)}\, , \\
    \widehat \npre(123, v)=\,\, & T_{(1)}\star T_{(2)}\star T_{(3)}=T_{(123)}-T_{(12),(3)}-T_{(13),(2)}\,, \\
    \widehat   \npre(1234, v)=&-T_{(1234)}+T_{(123),(4)}+T_{(14),(23)}+T_{(124),(3)}+T_{(12),(34)}-T_{(12),(3),(4)}\nn\\
      &-T_{(12),(4),(3)} -T_{(14),(2),(3)} +T_{(134),(2)}+T_{(13),(24)}-T_{(13),(2),(4)}\nn\\
      &-T_{(13),(4),(2)}-T_{(14),(3),(2)} \, .
\end{align}
In general, the fusion product is captured by the following formula, 
\begin{align}
\label{fusionone}
	 &T_{(1\tau_1),\ldots, (\tau_r )}\star T_{(j)}= \hskip-0.5cm  \sum_{\sigma\in \{(\tau_1),\ldots, (\tau_{r})\}\shuffle \{(j)\}}T_{(1\sigma_1),\ldots, (\sigma_{r+1})} -\sum_{i=1}^{r}T_{(1\tau_1),\ldots, (\tau_{i-1}),(\tau_{i}j),(\tau_{i+1}), \ldots,(\tau_{r})}\, .
\end{align}
It is important to stress that the fusion products, or more generally the algebra, are identical to that given in the references  \cite{Brandhuber:2021bsf,Chen:2022nei, Brandhuber:2022enp}, which was introduced to describe BCJ numerators of Yang-Mills theory (i.e. without higher-derivative corrections). 

Following \cite{Brandhuber:2021bsf,Chen:2022nei, Brandhuber:2022enp}, once $\widehat \npre(12\ldots n{-}2; v)$ is constructed in terms of the algebraic objects $T_{(1\tau_1),\ldots, (\tau_r )}$ using the fusion product rules given in \eqref{fusionone}, to obtain the BCJ numerators, we further introduce the evaluation maps which translate abstract generators into functions of  physical variables (such as momenta and polarisation vectors). The evaluation maps for the BCJ numerators of scattering amplitudes in Yang-Mills theory (with two heavy massive scalars) were written down and proved in \cite{Brandhuber:2021bsf}, and further extended to more general cases (with arbitrary numbers of scalars with general masses) \cite{Chen:2022nei}. As we emphasised previously that the fusion rules of the Hopf algebra we propose for the BCJ numerators of the higher-derivative theory \eqref{eq:Lag} is the same as that for the Yang-Mills theory \cite{Brandhuber:2021bsf,Chen:2022nei, Brandhuber:2022enp}, therefore the higher-derivative corrections can only arise in the evaluation maps, which we will discuss in the next subsection. 

\subsection{The evaluation maps}

We denote the evaluation maps of the generator  $T$ as $\langle T\rangle$ \cite{Brandhuber:2021bsf,Chen:2022nei, Brandhuber:2022enp}. To expose the pole structure of massive propagators, we propose that $\langle T\rangle$ takes the following form
\begin{align}\label{eq:T}
&\langle T_{(1\tau_1),(\tau_2),\ldots,(\tau_r)}\rangle:=\begin{tikzpicture}[baseline={([yshift=-0.8ex]current bounding box.center)}]\tikzstyle{every node}=[font=\small]   
   \begin{feynman}
    \vertex (a)[myblob]{};
     \vertex[left=0.8cm of a] (a0)[myblob]{};
     \vertex[right=0.8cm of a] (a2)[myblob]{};
      \vertex[right=0.8cm of a2] (a3)[myblob]{};
       \vertex[right=0.8cm of a3] (a4)[myblob]{};
       \vertex[above=0.8cm of a] (b1){$\tau_1~~~~$};
        \vertex[above=0.8cm of a2] (b2){$\tau_2$};
        \vertex[above=0.8cm of a3] (b3){$\cdots$};
         \vertex[above=0.8cm of a4] (b4){$\tau_r$};
         \vertex[above=0.8cm of a0] (b0){$1~~$};
       \vertex [above=0.8cm of a0](j1){$ $};
    \vertex [right=0.2cm of j1](j2){$ $};
    \vertex [right=0.6cm of j2](j3){$ $};
    \vertex [right=0.4cm of j3](j4){$ $};
    \vertex [right=0.8cm of j4](j5){$ $};
      \vertex [right=0.2cm of j5](j6){$ $};
    \vertex [right=0.6cm of j6](j7){$ $};
     \vertex [right=0.0cm of j7](j8){$ $};
    \vertex [right=0.8cm of j8](j9){$ $};
   	 \diagram*{(a)--[very thick](a0),(a)--[very thick](a2),(a2)--[very thick](a3), (a3)--[very thick](a4),(a0) -- [thick] (j1),(a) -- [thick] (j2),(a)--[thick](j3),(a2) -- [thick] (j4),(a2)--[thick](j5),(a4) -- [thick] (j8),(a4)--[thick](j9)};
    \end{feynman}  
  \end{tikzpicture}\nn\\
&={1\over n-2} {G_{1\tau_1}(v)\over v\mdot p_{1}}  {G_{\tau_2}(p_{\Theta(\tau_{2})})\over v\mdot p_{1\tau_1}}\, \cdots {G_{\tau_r}(p_{\Theta(\tau_{r})})\over v\mdot p_{1\tau_1\ldots \tau_{r-1}}} \, ,
\end{align}
where the lines between red blobs are massive propagators, and the external lines are gluons, which are grouped into sets, denoted as $1, \tau_1, \tau_2, \ldots, \tau_r$. The $\tau_i$s are ordered non-empty sets 
such that $\tau_1\cup\tau_2\cup\cdots\cup\tau_r=\{2,3, \ldots ,  n{-}2\}$ and $\tau_i\cap\tau_j=\emptyset$, i.e. they constitute a partition.   
The set $\Theta(\tau_i)$ consists of all indices to the left of $\tau_i$ and smaller than the first index in $\tau_i$:
\begin{align}
    \Theta(\tau_i)= (\{1\}\cup \tau_1 \cup \cdots \cup \tau_{i-1}) \cap \{1,\ldots, \tau_{i[1]}\} \, . 
\end{align}
  The expression on the second line of \eqref{eq:T} manifests the massive propagators in the large-mass limit, which are given by $v \cdot p_{1\tau_1 \ldots}$.

The BCJ numerator is determined from the pre-numerators via the following relation,  
\begin{align}
\mathcal{N}_{\mathrm{BCJ}}(1\alpha, v)= \mathcal{N}([1\alpha], v)\, , 
\end{align}
where $[\bullet]$ stands for the left-nested commutator, defined as
\begin{align}
[i_1 i_2 \cdots i_r]=[\ldots [[i_1,i_2],i_3],\ldots, i_r] \, . 
\end{align}
For instance, $\mathcal{N}_{\mathrm{BCJ}}(12, v)=\mathcal{N}(12, v)-\mathcal{N}(21, v)$. This can also be expressed as~\cite{Chen:2021chy}
\begin{align} 
\npre([1\ldots n{-}2], v)=\neco(1\ldots n-2)\circ \npre(1\ldots n{-}2, v)\, ,
\end{align}
 where the left-nested operator is defined as 
\begin{align}
\neco(i_1,\ldots, i_m)\equiv \Big[\mathbb{I}-\mathbb{P}_{(i_2i_1)}\Big]\Big[\mathbb{I}-\mathbb{P}_{(i_3i_2i_1)}\Big]\cdots \Big[ \mathbb{I}-\mathbb{P}_{(i_m\cdots i_2i_1)}\Big]\,
\end{align}
and $\mathbb{P}_{(i_m\cdots i_2 i_1)}$ denotes the cyclic permutations, i.e. $\mathbb{P}_{(i_2 i_1)}$ is $(i_2\rightarrow i_1, i_1\rightarrow i_2)$, $\mathbb{P}_{(i_3 i_2 i_1)}$ is $(i_3\rightarrow i_2, i_2\rightarrow i_1, i_1\rightarrow i_3)$.
Then $\mathcal{N}([1\alpha], v)$ is obtained through the evaluation map \eqref{eq:T}, 
\begin{align}
    \mathcal{N}(1\alpha, v)=\langle \widehat\npre(1\alpha,v) \rangle\, ,
\end{align}
with $\widehat\npre(1\alpha,v)$ defined by the fusion product \eqref{eq:Nhat}. 

The kinematic algebra and the structure \eqref{eq:T} together automatically induce a recursive relation of the pre-numerator 
\begin{align}\label{eq:recJ}
    \npre(1\alpha, v)&=(-1)^{n-3}\Big( { G_{1\alpha}(v)\over (n-2) v\mdot p_{1}} \Big) \nn\\
   &+\sum_{\tau_L\subset \alpha}(-1)^{|\tau_R|}\Big((n-2-|\tau_R|)
   {\npre(1\tau_L, v)  G_{\tau_R}(p_{\Theta_L(\tau_R)}) \over (n-2)v\mdot p_{1\tau_L}}\Big)\, ,
\end{align}
where again $\{1\tau_L\}\cup\{\tau_R\}=\{1\alpha\}=\{12\ldots n-2\}$ and $|\tau_R|$ denotes the number of gluons in $\tau_R$. It is straightforward to see that the number of terms of $\npre(1,2,\ldots,n{-}2;v)$ is nothing but the Fubini numbers~\cite{Brandhuber:2021bsf} 
\begin{align}
    \mathsf{F}_{n{-}3}=\sum_{i=0}^{n-4}\binom{n{-}3}{i} \mathsf F_{i}\, ,
\end{align}
where $\mathsf{F}_0=1,\mathsf{F}_1=1$.

\section{Evaluation maps with higher-derivative corrections}\label{KH}

In this section, we will study the evaluation maps in the expression \eqref{eq:T}, namely the $G$-function. We will constrain it using a variety of physical properties such as factorisation behaviour, power counting, gauge invariance, and crossing symmetry, and then propose our solution.

\subsection{Factorisation behaviour}

We first note, as for the Yang-Mills theory without higher-derivative corrections, for certain special cases, the generators are mapped to $0$. In particular, we have 
\begin{align}
\la T_{(1),\ldots}\ra =0 \, ,  \qquad \la T_{\ldots,(1\ldots),\ldots}\ra=0 \, . 
\end{align}
In general, the $G$ function is non-trivial. For a given ordered set of gluons, say $\tau$,  we propose it takes the following universal form, 
\begin{align}\label{eq:Gfun}
    &G_{\tau}(x)= \Big(x\mdot F_{\tau}\mdot v+\sum_{ \sigma_1 i\sigma_2 j  \sigma_3 =\tau}x\mdot F_{\sigma_1}\mdot p_{i} W(i\sigma_2j) p_{j}\mdot F_{\sigma_3}\mdot v\Big)\,,
\end{align}
where the first term $x\mdot F_{\tau}\mdot v$ is precisely the result of the heavy-mass effective field theory (HEFT) \cite{Brandhuber:2021bsf} without higher-derivative operators. $W(i\sigma_2 j)$ contains the information of higher-derivative corrections, and is given by
\begin{align}\label{Wfunc}
   & W(i\sigma j)=\alpha' \tr(F_{[i\sigma]}F_{j})
    +\alpha'^2\sum_{\rho \in [i\sigma]j}\sum_{\rho_1 j_2\rho_2 j_1={\rm cyc}(\rho)}
    p_{j_1}\mdot F_{\rho_1}\mdot p_{j_2} \tr(F_{[j_2\rho_2] }F_{j_1})\nn\\
 &+\alpha'^2\sum_{\sigma_1j_1\sigma_2j_2\sigma_3=\sigma}\tr(F_{[i\sigma_1] }F_{j_1})p_{j_1}\mdot F_{\sigma_2}\mdot p_{j_2}\tr(F_{[j_2\sigma_3]}F_j)\, .
\end{align}
The summation in \eqref{eq:Gfun} is over all the partitions of ordered $\tau$ in terms of three ordered subsets $\sigma_1, \{ i\sigma_2 j\},\sigma_3$ (including empty $\sigma_i$). For example,
\begin{align}
    G_{12}(x)&=x\mdot F_{12}\mdot v+x\mdot p_1 W(12) p_2\mdot v \\
     G_{123}(x)&=x\mdot F_{123}\mdot v+x\mdot p_1 W(123) p_2\mdot v+x\mdot F_1 \mdot p_2 W(23) p_3\mdot v+x \mdot p_1 W(12) p_2\mdot F_3\mdot v \, . \nn
\end{align}

Below we will motivate the structure of $G_{\tau}(x)$, especially from the factorisation behaviour on the massive poles, also which imposes strong constraints on the function $W(i\sigma_2 j)$.

First, the term $x\mdot F_\tau\mdot v$, corresponding to Yang-Mills interactions with no higher derivatives, is determined by mass dimension counting, gauge invariance, linearity in $v$ (except for the case $x=v$, when it is quadratic in $v$), and the ordering of gluons in the set $\tau$. Next, the term corresponding to the higher derivative corrections is the most natural way to generalise the original term  $x\mdot F_\tau\mdot v$ by including extra powers of momenta, in a way consistent with the ordering in $\tau$. Corrections of order $\alpha'$ require two extra powers of momenta, and the simplest way to add two momenta  is by inserting them inside the chain of $F_\tau$. This can be either adjacent to each other, $x\mdot F_{\sigma_1}\mdot p_X p_Y\mdot F_{\sigma_2}\mdot v$, with $\sigma_1 \sigma_2=\tau$, or more generally, $x\mdot F_{\sigma_1}\mdot p_X W(F_{\sigma_2})p_Y\mdot F_{\sigma_3}\mdot v$, with $\sigma_1\sigma_2 \sigma_3=\tau$. Finally, the labels $X$ and $Y$ should be related to  the location of the insertions. At order $\alpha'$ $W$ is only a function of field strengths, while at $\alpha'^2$ it can include another two powers of extra momenta. Comparing this general expectation to explicit low point examples, we indeed find they satisfy \eqref{eq:Gfun}.


We will now further show that $G_{\tau}(x)$ defined in~\eqref{eq:Gfun} gives the correction factorisation behaviours of the numerators on the heavy mass propagators,
    \begin{align}\label{eq:facBhv}
    \npre(1\ldots n{-}2, v)\rightarrow {(n-2-|i\tau_R|)|i\tau_R|\over n-2}{p_{\Theta_L(i\tau_R)}\mdot p_i\over v\mdot p_{1\tau_L}}\npre(1\tau_L, v) \npre(i\tau_R, v) \, .
    \end{align}
Let us begin with the four-point case. The pre-numerator takes the following form, 
\begin{align}\label{eq:npre12}
	\npre(12,v)=-{G_{12}(v)\over 2v\mdot p_1}=-{v\mdot F_{12}\mdot v+v\mdot p_1 W(12) p_2\mdot v \over  2v\mdot p_1}\, .
\end{align}
We see that the higher-derivative correction term has no pole, therefore the factorisation is identical to that of Yang-Mills theory without corrections \cite{Brandhuber:2021bsf}. 

Terms containing $W$'s in the  $G$-function become relevant for higher-point numerators. The first such example is the five-point pre-numerator, which is given as
\begin{align}
	\npre(123,v)&={1\over 3}\left( {G_{123}(v)\over v\mdot p_1}-{G_{12}(v)\over v\mdot p_1}{G_{3}(p_{12})\over v\mdot p_{12}}-{G_{13}(v)\over v\mdot p_1}{G_{2}(p_{1})\over v\mdot p_{13}} \right) \\
	&={v\mdot F_{123}\mdot v+v\mdot p_1 W(123) p_3\mdot v+v\mdot p_1 W(12) p_2\mdot F_3\mdot v+v\mdot F_1 \mdot p_2 W(23) p_3\mdot v \over  v\mdot p_1}\nn\\
	&-\frac{ \left(v\mdot F_{12}\mdot v+W(12) p_1\mdot v p_2\mdot v\right)p_{12}\mdot F_3\mdot v}{p_1\mdot v p_{12}\mdot v}-\frac{ \left(v\mdot F_{13}\mdot v+W(13) p_1\mdot v p_3\mdot v\right)p_1\mdot F_2\mdot v}{p_1\mdot v p_{13}\mdot v} \, . \nn
\end{align}
The factorisation behaviour can be sketched as below,  
\begin{align} \label{eq:pic1}
     \begin{tikzpicture}[baseline={([yshift=-0.4ex]current bounding box.center)}]\tikzstyle{every node}=[font=\small]
		\begin{feynman}
			\vertex (p1) {\(v\)};
			\vertex [right=0.8cm of p1] (b1) [dot]{};
			\vertex [right=0.5cm of b1] (b2) []{};
			\vertex [right=0.5cm of b2] (b3) [HV]{H};
   \vertex [above=1.2cm of b1] (u1) []{$1$}; \vertex [above=1.2cm of b3] (u3) []{};\vertex [left=0.5cm of u3] (u3a) []{$2$};\vertex [right=0.5cm of u3] (u3b) []{$3$};
			\vertex [right=0.8cm of b3](p4){};
			\vertex [right=0.5cm of b1] (cut1);
			\vertex [above=0.2cm of cut1] (cut1u);
			\vertex [below=0.2cm of cut1] (cut1b);
			\vertex [right=0.25cm of b2] (cut2);
			\vertex [above=0.2cm of cut2] (cut2u);
			\vertex [below=0.2cm of cut2] (cut2b);
			\diagram* {
			(b1)--[gluon, thick](u1), (b3)-- [gluon, thick] (u3b),(b3)-- [gluon, thick] (u3a), (p1) -- [thick] (b1)-- [thick] (b3)-- [thick] (p4), (cut1u)--[ red,thick] (cut1b),
			};
		\end{feynman}
  \end{tikzpicture} &&
  \begin{tikzpicture}[baseline={([yshift=-0.4ex]current bounding box.center)}]\tikzstyle{every node}=[font=\small]
		\begin{feynman}
			\vertex (p1) {\(v\)};
			\vertex [right=0.8cm of p1] (b1) [HV]{H};
			\vertex [right=0.5cm of b1] (b2) []{};
			\vertex [right=0.5cm of b2] (b3) [dot]{};
   \vertex [above =1.2cm of p1] (u0) []{};\vertex [right =0.2cm of u0] (u1) []{$1$}; \vertex [above=1.2cm of b3] (u3) []{};\vertex [left=0.5cm of u3] (u3a) []{$2$};\vertex [right=0.0cm of u3] (u3b) []{$3$};
			\vertex [right=0.8cm of b3](p4){};
			\vertex [right=0.5cm of b1] (cut1);
			\vertex [above=0.2cm of cut1] (cut1u);
			\vertex [below=0.2cm of cut1] (cut1b);
			\vertex [right=0.25cm of b2] (cut2);
			\vertex [above=0.2cm of cut2] (cut2u);
			\vertex [below=0.2cm of cut2] (cut2b);
			\diagram* {
			(b1)--[gluon, thick](u1), (b3)-- [gluon, thick] (u3b),(b1)-- [gluon, thick] (u3a), (p1) -- [thick] (b1)-- [thick] (b3)-- [thick] (p4), (cut1u)--[ red,thick] (cut1b),
			};
		\end{feynman}
  \end{tikzpicture} && 
   \begin{tikzpicture}[baseline={([yshift=-0.4ex]current bounding box.center)}]\tikzstyle{every node}=[font=\small]
		\begin{feynman}
			\vertex (p1) {\(v\)};
			\vertex [right=0.8cm of p1] (b1) [HV]{H};
			\vertex [right=0.5cm of b1] (b2) []{};
			\vertex [right=0.5cm of b2] (b3) [dot]{};
   \vertex [above =1.2cm of p1] (u0) []{};\vertex [right =0.2cm of u0] (u1) []{$1$}; \vertex [above=1.2cm of b3] (u3) []{};\vertex [left=0.5cm of u3] (u3a) []{$3$};\vertex [right=0.0cm of u3] (u3b) []{$2$};
			\vertex [right=0.8cm of b3](p4){};
			\vertex [right=0.5cm of b1] (cut1);
			\vertex [above=0.2cm of cut1] (cut1u);
			\vertex [below=0.2cm of cut1] (cut1b);
			\vertex [right=0.25cm of b2] (cut2);
			\vertex [above=0.2cm of cut2] (cut2u);
			\vertex [below=0.2cm of cut2] (cut2b);
			\diagram* {
			(b1)--[gluon, thick](u1), (b3)-- [gluon, thick] (u3b),(b1)-- [gluon, thick] (u3a), (p1) -- [thick] (b1)-- [thick] (b3)-- [thick] (p4), (cut1u)--[ red,thick] (cut1b),
			};
		\end{feynman}
  \end{tikzpicture}.
\end{align}
Since the factorisation of the leading-order terms in $\alpha'$ (i.e. terms without $W$) has already been shown in \cite{Brandhuber:2021bsf}, we will only focus on the terms beyond the leading order, It is straightforward to see that from the explicit expression of $\npre(123,v)$, at the three poles, $p_1\mdot v \to 0, p_{12}\mdot v \to 0$, and $p_{123}\mdot v \to 0$ (as shown in \eqref{eq:pic1}), $\npre(123,v)$ factorises as  
\begin{align}
    \Big({p_1\mdot p_2\over p_1\mdot v}\Big)\Big(v\mdot\veps_1\Big)\Big(-{p_2\mdot v  W(23) p_3\mdot v\over p_2\mdot v} \Big)\, , &&  \Big({p_{12}\mdot p_3\over p_{12}\mdot v}\Big)\Big(-{p_1\mdot v  W(12) p_2\mdot v\over p_{1}\mdot v} \Big)\Big(v\mdot\veps_3\Big) \, ,\nn\\
      \Big({p_{1}\mdot p_3\over p_{13}\mdot v}\Big)\Big(-{p_1\mdot v  W(13) p_3\mdot v\over p_1\mdot v} \Big)\Big(v\mdot\veps_2\Big) \, . &&
\end{align}
The above expressions agree with the general factorisation behaviour \eqref{eq:facBhv}. Once again importantly the term $W(123)$ does not contribute in the factorisation limit. The factorisation behaviour of six-point numerators can be analysed in a similar fashion. The explicit expression for the six-point numerator is given in \eqref{eq:six-num} of the appendix \ref{sec:sixpts}.  The factorisation behaviour on most of the heavy massive channels is similar.  The only non-trivial one is on the pole $1\over p_1\mdot v$, for which we find (for the terms with $\alpha'$ corrections)
\begin{align}
  {p_1 \mdot p_2 v \mdot \veps _1 \over p_1\mdot v} &\Big( W(2 3)  p_3 \mdot F_4 \mdot v-W(2 3 4)  p_{2 3} \mdot v \\
  &-\frac{W(2 3)  p_3 \mdot v  p_{2 3} \mdot F_4 \mdot v}{p_{2 3} \mdot v}-\frac{W(3 4)  p_{2 3} \mdot v v \mdot F_2 \mdot p_3}{p_2 \mdot v}-\frac{W(2 4)  p_{2 3} \mdot v p_2 \mdot F_3 \mdot v}{p_3 \mdot v}\Big) \, . \nn 
\end{align}
The result again agrees with the factorisation behaviour  \eqref{eq:facBhv}. These examples show that the evaluation maps \eqref{eq:T} and \eqref{eq:Gfun} are sufficient to ensure the factorisation behaviours even without knowing the explicit form of the $W$-function, and the analysis  further confirms that the function $G_{\tau}(x)$ has the correct form.

\subsection{Crossing symmetry}

In the following we will constrain the function $W(i\sigma_2 j)$. The  general properties of $W(i\sigma_2 j)$ are 
\begin{itemize}
	\item It is manifestly gauge invariant without any poles, therefore it should composed of the product of the factor $\tr(F_\sigma)$ or $p_X\mdot F_\sigma\mdot p_Y$
	\item By power counting, it should have mass dimension $|\sigma_2|$, where $|\sigma_2|$ denotes the number of the gluon labels in $\sigma_2$.

 \item It is constrained by the full crossing symmetry of the BCJ numerators. 
\end{itemize}

We will now analyse in details the constraints on $W(i\sigma_2 j)$ due to the crossing symmetry on the BCJ numerator, as in the case of no higher-derivative corrections \cite{Brandhuber:2021bsf, Brandhuber:2022enp}. The crossing symmetry results in  a novel relation between the pre-numerator and BCJ numerator 
\begin{align}\label{eq:preNtoN}
\neco(1\ldots n-2)\circ \npre(1\ldots n{-}2, v)= (n{-}2)\npre(1\ldots n{-}2, v)\, ,
\end{align}
and we will refer this as the crossing symmetry relation.

At four points, the crossing symmetry relation requires that 
\begin{align}
    \neco(1,2)\circ \npre(12,v)=2 \npre(12,v)\, .
\end{align}
From  \eqref{eq:npre12}, this implies that $W(12)$ is a symmetric function under the swapping of label $1$ and $2$.  The relative coefficient between these terms may be fixed using the factorisation property of the amplitude for the massless poles, and we find the $W(12)$ can only be a linear combination of the following two terms 
\begin{align}
	\alpha' \tr(F_{12}),  && \alpha'^2 p_{12}^2\tr (F_{12}) \, , 
\end{align}
for the first two orders in the $\alpha'$-expansion.\footnote{Other possibilities such as $\alpha'^2 p_2\mdot F_1\mdot F_2\mdot p_1$ are not independent. For example, $\alpha'^2 p_2\mdot F_1\mdot F_2\mdot p_1$ is  proportional to $\alpha'^2 p_{12}^2\tr (F_{12})$, so we do not include it.} The coefficients of these terms may be fixed using the factorisation property of the amplitude for the massless poles, which turn out to be simply $1$, so we find
\begin{align}
	W(12)=\alpha' \tr(F_{12})+\alpha'^2 p_{12}^2\tr (F_{12}) \, .  
\end{align}
At five points, the crossing symmetry relation is 
\begin{align}
    \neco(1,2,3)\circ \npre(123,v)&=\npre(123,v)-\npre(213,v)-\npre(312,v)+\npre(321,v)\nn\\
    &=3 \npre(123,v) \, .
\end{align}
After using the crossing symmetry property for the $W(i_1 i_2)$, the above relation implies the following relation on  $W(123)$, 
\begin{align} \label{eq:LW123}
    \neco(1,2,3)\circ (W(123)p_3\mdot v) =3 W(123)p_3\mdot v \, ,
\end{align}
under the on-shell condition $p_{123}\mdot v=0$.
Using the explicit definition of the operator $\neco(1,2,3)$, we find \eqref{eq:LW123} yields following conditions 
\begin{align}\label{eq:W123}
    2W(123)+W(213)+W(321)=0\, , && W(312)+W(321)=0 \, .
\end{align}
From  power counting as well as manifest gauge invariance, we find $W(123)$ should be consisted of the following terms,  
\begin{align}
\alpha':& \qquad \tr(F_{123}) \, , \\
\alpha'^2:&\qquad 
 p_{1}\mdot p_2 \tr(F_{123})\, , \quad p_{1}\mdot p_3 \tr(F_{123})\, , \quad p_{2}\mdot p_3 \tr(F_{123})\, ,  \nn\\
   & \qquad   p_{1}\mdot F_3\mdot p_2 \tr(F_{12})\, , \quad p_{1}\mdot F_2\mdot p_3 \tr(F_{13})\, , \quad p_{2}\mdot F_1\mdot p_3 \tr(F_{23}) \, . 
\end{align}
We then find that there are only three solutions, which are consistent with the crossing symmetric constraints \eqref{eq:W123},\footnote{We write terms in this way to manifest the symmetries. For example, in fact $\tr(F_{213})=-\tr(F_{123})$, so one may write $\tr(F_{123})-\tr(F_{213})=2\, \tr(F_{123})$.} 
\begin{align} \label{eq:5pts-terms}
 (1):~~  & \tr(F_{123})-\tr(F_{213})\,,\\
  (2):~~ &  p_{123}^2 (\tr(F_{123})-\tr(F_{213}))\, ,\nn\\
  (3):~~ &   \tr(F_{12}) p_2\mdot F_3\mdot p_1+\tr(F_{23}) p_3\mdot F_1\mdot p_2+\tr(F_{31}) p_1\mdot F_2\mdot p_3\nn\\
  &-\tr(F_{13}) p_3\mdot F_2\mdot p_1-\tr(F_{21}) p_1\mdot F_3\mdot p_2-\tr(F_{32}) p_2\mdot F_1\mdot p_3 \, . \nn
\end{align}
Interestingly, all the three solutions have another extra common crossing symmetry under reversing the indices $\{1,2,3\}\rightarrow \{3,2,1\}$. So we also have 
\begin{align}\label{eq:W123Relation2}
    W(123)=-W(321).
\end{align}
Under this relation, the first relation in  \eqref{eq:W123} is not an independent one anymore and  the independent relations are 
\begin{align}\label{eq:W123idr}
   W(312)+W(321)=0\, ,  && W(123)=-W(321)\, .
\end{align}
By further comparing the factorisations on massless poles (or even the amplitude), we fix all the coefficients for the terms in \eqref{eq:5pts-terms}. Once again, we find all the coefficients are simply $1$, and so
\begin{align}\label{eq:W123Res}
    W(123)=\alpha' \tr(F_{[12]}F_{3})
    +\alpha'^2\sum_{\rho \in [12]3}\sum_{\rho_1 j_2\rho_2 j_1={\rm cyc}(\rho)}
    p_{j_1}\mdot F_{\rho_1}\mdot p_{j_2} \tr(F_{[j_2\rho_2] }F_{j_1}) \, ,
\end{align}
where $F_{[12]} = F_1\mdot F_2 - F_2\mdot F_1$. 
Note when $\rho_1$ is an empty set, the above expression leads to the term $(2)$ in \eqref{eq:5pts-terms}. 

 At six points, the crossing symmetric condition implies
\begin{align}
  &\neco(1,2,3,4)\circ \Bigg(\Big(W(1234)-  W(12) W(34) p_2\mdot p_3\Big) p_4\mdot v\Bigg)\nn\\
  &= 4\Big(W(1234)-  W(12) W(34) p_2\mdot p_3\Big) p_4\mdot v \, ,
\end{align}
where we have used the on-shell condition $p_{1234}\mdot v=0$ and the symmetry property of $W(i_1 i_2)$ and $W(i_1 i_2 i_3)$.  
For convenience, we define 
\begin{align}
W'(1234)= W(1234)-  W(12) W(34) p_2\mdot p_3 \, ,
\end{align} 
where the second term, which we subtract, can be viewed as a product of lower-point order-$\alpha'$ terms. 
Then we have the relations 
\begin{align}\label{eq:W1234}
   & W'(4312)+W'(4321)=0 \, ,  \\
& W'(4123)-W'(4213)-W'(4321)=0 \, ,  \\
   & 3 W'(1234)+W'(2134)+W'(3124)-W'(3214)-W'(4321)=0 \, .
\end{align}
Below are the terms that are consistent with power counting and manifest gauge invariance at six points: at the $\alpha'$ order, we have,\footnote{We have omitted double-trace terms (i.e. terms proportional to $\tr(F_{12})\tr(F_{34})$) in writing down possible terms for $W'(1234)$, since they are already contained in $ W(12) W(34)$, and furthermore they are inconsistent with the relations \eqref{eq:W1234}.} 
\begin{align}
    \tr(F_{1234}), && \tr(F_{2134}),
\end{align}
and for the $\alpha'^2$ order
\begin{align}
   p_{i}\mdot p_j \tr(F_{1234}), && p_{i}\mdot p_j\tr(F_{2134}), && p_{i}\mdot F_4\mdot p_j \tr(F_{123}),  
   \nn\\ 
 p_{i}\mdot F_3\mdot p_j \tr(F_{124}) \, , && p_{i}\mdot F_2\mdot p_j \tr(F_{134})\, ,&& p_{i}\mdot F_1\mdot p_j \tr(F_{234}) \, ,  \nn\\ 
       p_{i}\mdot F_{34}\mdot p_j \tr(F_{12}) \, , &&  p_{i}\mdot F_{24}\mdot p_j \tr(F_{13}), && p_{i}\mdot F_{23}\mdot p_j \tr(F_{14}) \, ,  \nn\\
      p_{i}\mdot F_{14}\mdot p_j \tr(F_{23})\, , &&  p_{i}\mdot F_{13}\mdot p_j \tr(F_{24})\, , &&  p_{i}\mdot F_{12}\mdot p_j \tr(F_{34})\, . && 
\end{align}
Then the solutions of \eqref{eq:W1234} are 
\begin{align}
(1):~& \tr(F_{1234})- \tr(F_{2134})- \tr(F_{3124})+\tr(F_{3214}) \, ,\\
(2):~& p_{1234}^2\tr(F_{1234})- p_{1234}^2\tr(F_{2134})- p_{1234}^2\tr(F_{3124})+p_{1234}^2\tr(F_{3214}) \, , \nn \\
(3):~&\tr(F_{[12]4}) p_1\mdot F_3\mdot p_4-\tr(F_{[13]2}) p_1\mdot F_4\mdot p_2+\tr(F_{[13]4}) p_1\mdot F_2\mdot p_4-\tr(F_{[14]2}) p_1\mdot F_3\mdot p_2\nn\\
&+\tr(F_{[21]3}) p_2\mdot F_4\mdot p_3+\tr(F_{[23]1}) p_1\mdot F_4\mdot p_2+\tr(F_{[24]1}) p_1\mdot F_3\mdot p_2+\tr(F_{[24]3}) p_2\mdot F_1\mdot p_3\nn\\
&-\tr(F_{[31]2}) p_2\mdot F_4\mdot p_3-\tr(F_{[31]4}) p_3\mdot F_2\mdot p_4-\tr(F_{[32]4}) p_3\mdot F_1\mdot p_4-\tr(F_{[34]2}) p_2\mdot F_1\mdot p_3\nn\\
&+\tr(F_{[41]3}) p_3\mdot F_2\mdot p_4-\tr(F_{[42]1}) p_1\mdot F_3\mdot p_4+\tr(F_{[42]3}) p_3\mdot F_1\mdot p_4-\tr(F_{[43]1}) p_1\mdot F_2\mdot p_4\, ,\nn \\
(4):~&\tr(F_{12}) p_1\mdot F_{43}\mdot p_2-\tr(F_{13}) p_1\mdot F_{24}\mdot p_3+\tr(F_{13}) p_1\mdot F_{42}\mdot p_3-\tr(F_{14}) p_1\mdot F_{23}\mdot p_4\nn\\
&-\tr(F_{21}) p_1\mdot F_{34}\mdot p_2-\tr(F_{23}) p_2\mdot F_{14}\mdot p_3-\tr(F_{24}) p_2\mdot F_{13}\mdot p_4+\tr(F_{24}) p_2\mdot F_{31}\mdot p_4\nn\\
&-\tr(F_{31}) p_1\mdot F_{24}\mdot p_3+\tr(F_{31}) p_1\mdot F_{42}\mdot p_3+\tr(F_{32}) p_2\mdot F_{41}\mdot p_3+\tr(F_{34}) p_3\mdot F_{21}\mdot p_4\nn\\
&+\tr(F_{41}) p_1\mdot F_{32}\mdot p_4-\tr(F_{42}) p_2\mdot F_{13}\mdot p_4+\tr(F_{42}) p_2\mdot F_{31}\mdot p_4-\tr(F_{43}) p_3\mdot F_{12}\mdot p_4 \, , \nn\\
\cdots~ &\text{three other solutions}\cdots \, . \nn
\end{align}
The coefficients of these terms are fixed by massless poles (or directly the amplitude). It turns out, the terms of $(1-4)$ that are shown in the above have coefficient $1$, while ``three other solutions" that we did not show explicitly have vanishing coefficients. In conclusion, we find, 
\begin{align}\label{eq:W1234Res1}
    W'(1234)&=\alpha' \tr(F_{[123]}F_4)+\alpha'^2p_{1234}^2\tr(F_{[123]}F_4)\nn\\
    &+\alpha'^2\sum_{\rho=[123]4}\Big(p_{\rho_4}\mdot F_{\rho_1}\mdot p_{\rho_2}\tr(F_{[\rho_2\rho_3]}F_{\rho_4})+p_{\rho_1}\mdot F_{\rho_2}\mdot p_{\rho_3}\tr(F_{[\rho_3\rho_4]}F_{\rho_1})\nn\\
    &+p_{\rho_2}\mdot F_{\rho_3}\mdot p_{\rho_4}\tr(F_{[\rho_4\rho_1]}F_{\rho_2})+p_{\rho_3}\mdot F_{\rho_4}\mdot p_{\rho_1}\tr(F_{[\rho_1\rho_2]}F_{\rho_3})\Big)\nn\\
    &+\alpha'^2\sum_{\rho=[123]4}\Big(p_{\rho_4}\mdot F_{\rho_1\rho_2}\mdot p_{\rho_3}\tr(F_{\rho_3}F_{\rho_4})+p_{\rho_1}\mdot F_{\rho_2\rho_3}\mdot p_{\rho_4}\tr(F_{\rho_4}F_{\rho_1})\nn\\
    &+p_{\rho_2}\mdot F_{\rho_3\rho_4}\mdot p_{\rho_1}\tr(F_{\rho_1}F_{\rho_2})+p_{\rho_3}\mdot F_{\rho_4\rho_1}\mdot p_{\rho_2}\tr(F_{\rho_2}F_{\rho_3})\Big)\, ,
\end{align}
which can be recast nicely as
\begin{align}\label{eq:W1234Res}
    W'(1234)=\alpha' \tr(F_{[123]}F_{4})
    +\alpha'^2\sum_{\rho \in [123]4}\sum_{\rho_1 j_2\rho_2 j_1={\rm cyc}(\rho)}
    p_{j_1}\mdot F_{\rho_1}\mdot p_{j_2} \tr(F_{[j_2\rho_2] }F_{j_1}) \, , 
    \end{align}
where $\sum_{\rho=[123]4}f(\rho)\equiv f(1234)-f(2134)-f(3124)+f(3214)$ and $F_{[123]}\equiv F_{123}-F_{213}-F_{312}+F_{321}$. 

Interestingly, from the explicit expression of $W'(1234)$, we note it obeys an extra relation  by reversing the indices
\begin{align}
W'(1234)-W'(4321)=0 \, . 
\end{align}
Combined with this additional relation, the relations in \eqref{eq:W1234} for $W'$-function become
\begin{align}
    W'(4312)+W'(4321)=0\, , &&
 W'(4123)-W'(4213)-W'(4321)=0\, , \nn\\
  W'(1234)-W'(4321)=0\, .
\end{align}
From the $W'$, we get the form of $W$ function 
\begin{align}
    W(1234)= W'(1234)+W(12)p_2\mdot p_3 W(34)\, .
\end{align}
The first term can be understood as the primary contact terms which are constrained from the simple crossing symmetry. The contact terms contain two factors of $W$ functions arise from lower-point BCJ numerators, which can be viewed as the descendant contact terms. 

Let us now consider the general case. Up to the $\alpha'^2$ order, we find the $W(12\ldots r)$ function has the following structure
\begin{align}
    W(12\ldots r)\Big|_{\alpha'^2}=\Big(W'(12\ldots r)+\!\! \sum_{2\leq i<j\leq r-1} W(12\ldots i)p_i\mdot F_{i+1 \ldots j-1}\mdot p_j W(j\ldots r)\Big)\Big|_{\alpha'^2}\, ,
\end{align}
where the contact terms with two $W$-functions are constructed to preserve the ordering of gluons. The primary contact term (i.e. $W'(12\ldots r)$) satisfies the crossing symmetry
\begin{align} \label{eq:Wprime}
  \neco(1,2,\ldots,r)\circ \Big( W'(12\ldots r) \, p_r\mdot v\Big)= r \, W'(12\ldots r)  \, p_r\mdot v\, ,
\end{align}
valid under the on-shell condition $p_{12\ldots r}\mdot v=0$. 
This leads to the following symmetry properties of the $W'$-function 
\begin{align}
& \sum_{\sigma\in [\rho]}W'(\sigma i_{r}) - (-1)^{|\rho|} W'(i_{r} \rho^{\rm rev}) =r\, W'(i_1i_2\ldots i_r) \, , \\
 &   \sum_{\sigma\in [\rho]}W'(i_r\ldots i_{r-j-1}\sigma i_{r-j}) - (-1)^{|\rho|} W'(i_{r}\ldots i_{r-j-1} i_{r-j} \rho^{\rm rev}) =0 \, ,
\end{align}
where $[\rho]$ denotes left-nested commutators,  and $|\rho|$ is the size of $\rho$ and $\rho^{\rm rev}$ denotes the reversing of $\rho$. The second relation is valid for $j\in \{2,3,\ldots,r-1\}$. 

Following the patterns of five- and six-point examples, given in \eqref{eq:W123Res} and \eqref{eq:W1234Res}, respectively, it is straightforward to show that the following general solution is consistent with \eqref{eq:Wprime}, 
\begin{align} \label{eq:Wprimsol}
    W'(123\ldots r)=\alpha' \tr(F_{[1\ldots r-1]}F_{r})
    +\alpha'^2\sum_{\rho \in [1\ldots r-1]r}\sum_{\rho_1 j_2\rho_2 j_1={\rm cyc}(\rho)}
    p_{j_1}\mdot F_{\rho_1}\mdot p_{j_2} \tr(F_{[j_2\rho_2] }F_{j_1}) \, .
\end{align}
Once again, the explicit expression for $  W'(123\ldots r)$ given in \eqref{eq:Wprimsol} has an extra reversing symmetry
\begin{align}
   W'(\rho)-(-1)^{|\rho|} W'(\rho^{\rm rev})=0 \, .
\end{align}

Combining all the discussion above, we obtain  a closed formula for the $W$-function 
\begin{align}
   & W(i\sigma j)=\alpha' \tr(F_{[i\sigma]}F_{j})
    +\alpha'^2\sum_{\rho \in [i\sigma]j}\sum_{\rho_1 j_2\rho_2 j_1={\rm cyc}(\rho)}
    p_{j_1}\mdot F_{\rho_1}\mdot p_{j_2} \tr(F_{[j_2\rho_2] }F_{j_1})\nn\\
 &+\alpha'^2\sum_{\sigma_1j_1\sigma_2j_2\sigma_3=\sigma}\tr(F_{[i\sigma_1] }F_{j_1})p_{j_1}\mdot F_{\sigma_2}\mdot p_{j_2}\tr(F_{[j_2\sigma_3]}F_j)\, .
\end{align}
We have further checked explicitly the correctness of the BCJ numerators up to eight points against the amplitudes computed from the action \eqref{eq:Lag} using Berends-Giele recursion~\cite{Garozzo:2018uzj}. This completes our proposal of the kinematic Hopf algebra for the BCJ numerators with higher-derivative corrections, $\alpha' F^3$ and $\alpha'^2 F^4$. 

The leading correction (the term proportional to $\alpha'$) is due to the $F^3$ term in the action $S_{\mathrm{YM}+\alpha'F^3+\alpha'^2F^4}$ as defined in \eqref{eq:Lag}. By power counting and the gauge invariance condition, the linear $\alpha'$ term can only be the trace of the strength tensors. The particular combination in the nested commutator is related to the full crossing symmetry for the BCJ numerators.  The subleading $\alpha'^2$ corrections contain two distinct structures.  This is indeed expected, since this order receives contributions from the $\alpha'^2F^4$ operator, and also from an operator of the type $(\alpha' F^3)^2$.

\subsection{Summary and pure gluon amplitudes}

Here we briefly summarise our final results of constructing BCJ numerators using the quasi-Hopf algebra: 
\begin{align}
\mathcal{N}_{\mathrm{BCJ}}(1\alpha, v)= \mathcal{N}([1\alpha], v)\, , 
\end{align}
where $[ \bullet ]$ denotes left-nested commutators, and $\mathcal{N}([1\alpha], v)$ is obtained through the evaluation map, 
\begin{align}
    \mathcal{N}(1\alpha, v)=\langle \widehat\npre(1\alpha,v) \rangle\, .
\end{align}
The algebraic numerator $\widehat\npre$ is defined as, 
\begin{align}
 \widehat\npre(12\ldots n{-}2, v)=T_{(1)}\star T_{(2)}\cdots \star T_{(n{-}2)} \, ,
 \end{align}
with fusion product rules given in \eqref{fusionone}. The evaluation map is defined as
\begin{align}\label{eq:T2}
&\langle T_{(1\tau_1),(\tau_2),\ldots,(\tau_r)}\rangle={1\over n-2} {G_{1\tau_1}(v)\over v\mdot p_{1}}  {G_{\tau_2}(p_{\Theta(\tau_{2})})\over v\mdot p_{1\tau_1}}\, \cdots {G_{\tau_r}(p_{\Theta(\tau_{r})})\over v\mdot p_{1\tau_1\ldots \tau_{r-1}}} \, ,
\end{align}
where
\begin{align}\label{eq:Gfun2}
    &G_{\tau}(x)= \Big(x\mdot F_{\tau}\mdot v+\sum_{\sigma_1  i\sigma_2 j  \sigma_3=\tau}x\mdot F_{\sigma_1}\mdot p_{i} W(i\sigma_2j) p_{j}\mdot F_{\sigma_3}\mdot v\Big)\,,
\end{align}
and the $W(i\sigma j)$ term gives $\alpha'$ corrections and is given by
\begin{align}\label{Wfunct}
   & W(i\sigma j)=\alpha' \tr(F_{[i\sigma]}F_{j})
    +\alpha'^2\sum_{\rho \in [i\sigma]j}\sum_{\rho_1 j_2\rho_2 j_1={\rm cyc}(\rho)}
    p_{j_1}\mdot F_{\rho_1}\mdot p_{j_2} \tr(F_{[j_2\rho_2] }F_{j_1})\nn\\
 &+\alpha'^2\sum_{\sigma_1j_1\sigma_2j_2\sigma_3=\sigma}\tr(F_{[i\sigma_1] }F_{j_1})p_{j_1}\mdot F_{\sigma_2}\mdot p_{j_2}\tr(F_{[j_2\sigma_3]}F_j)\, .
\end{align}
Finally, as argued in \cite{Brandhuber:2021bsf},  by a simple factorisation limit as shown below, 
\begin{align} \label{eq:puregluon}
    \begin{tikzpicture}[baseline={([yshift=-1.5ex]current bounding box.center)}]\tikzstyle{every node}=[font=\small]    
   \begin{feynman}
    \vertex (a)[myblob]{};
    \vertex [above=0.3cm of a](aa){\red{$\bm\times$}};
     \vertex [above=0.5cm of a](b)[dot]{};
     \vertex [left=0.6cm of b](c);
     \vertex [left=0.22cm of b](c23);
     \vertex [above=0.13cm of c23](v23)[dot]{};
    \vertex [above=.4cm of c](j1){$1$};
    \vertex [right=.7cm of j1](j2){$2\cdots$};
    \vertex [right=0.5cm of j2](j3){$~~n{-}2$};
   	 \diagram*{(a) -- [thick] (b),(b) -- [thick] (j1),(v23) -- [thick] (j2),(b)--[thick](j3)};
    \end{feynman}  
  \end{tikzpicture}& \begin{tikzpicture}[baseline={([yshift=2.0ex]current bounding box.center)}]\tikzstyle{every node}=[font=\small]    
   \begin{feynman}
    \vertex (a){$\longrightarrow$};
   	 \diagram*{};
    \end{feynman}  
  \end{tikzpicture}~~~
  \begin{tikzpicture}[baseline={([yshift=-0.8ex]current bounding box.center)}]\tikzstyle{every node}=[font=\small]    
   \begin{feynman}
    \vertex (a){$n{-}1$};
     \vertex [above=0.6cm of a](b)[dot]{};
     \vertex [left=0.6cm of b](c);
     \vertex [left=0.22cm of b](c23);
     \vertex [above=0.13cm of c23](v23)[dot]{};
    \vertex [above=.4cm of c](j1){$1$};
    \vertex [right=.7cm of j1](j2){$2\cdots$};
    \vertex [right=0.5cm of j2](j3){$~~n{-}2$};
   	 \diagram*{(a) -- [thick] (b),(b) -- [thick] (j1),(v23) -- [thick] (j2),(b)--[thick](j3)};
    \end{feynman}  
  \end{tikzpicture}\, ,
\end{align}
one can simply decouple the heavy-massive scalars and obtain the BCJ numerators of pure gluons from $\mathcal{N}([1\alpha], v)$. Explicitly, the BCJ numerators of $(n{-}1)$ gluons can be deduced from the numerators with scalars as a limit, 
\begin{align} \label{eq:gluon}
\mathcal{N}^{\rm gluon}([1\ldots n{-}1]) &= \mathcal{N}([1\ldots n{-}2], v)\big{|}^{v \rightarrow \epsilon_{n{-}1}}_{ p_{1\ldots n{-}2}^2\rightarrow 0} \, ,
\end{align}
where $p_{1\ldots n{-}2}^2\rightarrow 0$ is used to impose the on-shell condition for the $(n{-}1)$-th gluon that arises through the factorisation limit, as shown in \eqref{eq:puregluon}. 

Once we obtain the BCJ numerators, using the inverse of \eqref{eq:KLT}, the gluon and graviton amplitudes (with two heavy particles) can be constructed as 
\begin{align} \label{eq:KLT2}
    A(1\alpha,v)&=\sum_{\beta\in S_{n-3}} \mathrm{m}(1\alpha, 1\beta)\,\npre([1\beta],v)\,, \\
\label{eq:KLT3}     M(1\alpha,v)&= \sum_{\alpha, \beta\in S_{n-3}}\npre([1\alpha],v)\,\mathrm{m}(1\alpha, 1\beta)\,\npre([1\beta],v)\,, 
\end{align}
where $\mathrm{m}(1\alpha, 1\beta)$ is the propagator matrix \cite{Vaman:2010ez} which equals to the inverse of the KLT matrix $\mathcal{S}$ that appears in \eqref{eq:KLT}. The summation is over all the permutation of the $n-3$ massless particles. The KLT formulas \eqref{eq:KLT2} and \eqref{eq:KLT3} also apply to pure-gluon and pure-graviton amplitudes, which can be achieved by imposing the condition \eqref{eq:gluon} on the formulas. 
 

\section{Conclusions and outlook}\label{conc}

We have demonstrated that higher-derivative operators that are compatible with colour-kinematics also admit a Hopf algebra representation. To be specific, we have considered higher-derivative operators of the form $\alpha' F^3$ and $\alpha'^2 F^4$ that were first studied in \cite{Broedel:2012rc} in the context of colour-kinematics duality. Understanding the underlying algebraic structure leads to a very compact formula for the BCJ numerators of any number of particles. Our results strongly suggest that the Hopf algebra may in fact underlie  all instances of the colour-kinematic duality, which is known to apply to a wide range of theories. Our construction also hints on the possibility of observing the kinematic algebra at the Lagrangian level. In particular, the structure of the evaluation maps suggests that there is a direct connection between higher-derivative operators and terms in the evaluation maps. 

There are two other natural research directions we could investigate in the future. The first one is to understand the evaluation maps for the theories with higher-derivative operators beyond the $\alpha'^2$ order. Particularly, our ansatz for the $G$-function may be valid even in the higher orders of $\alpha'$, and the structure of the $W$-function is very suggestive hinting on possible higher-order generalisations. Such generalisations could then complement studies on higher derivative operators from positivity conditions on EFT's that satisfy double copy (such as gravity), or monodromy relations (in the case of string theory) \cite{Camanho:2014apa,Huang:2020nqy,Bern:2021ppb,Caron-Huot:2022ugt,Chiang:2022jep,Chiang:2023quf,Berman:2023jys,Carrasco:2023wib}. It would be fascinating if the Hopf algebra reveals any further structure or relations between these operators.  We may also obtain BCJ numerators of other theories, in particular for scalars, from our results using the transmutation procedure~\cite{Cheung:2017ems} and compare with other recent bootstrap approaches (e.g. \cite{Chen:2022shl,Pavao:2022kog,Carrasco:2022sck,Chen:2023dcx,Li:2023wdm,Brown:2023srz}).

The second one is to generalise the construction beyond purely kinematic numerators following~\cite{Carrasco:2019yyn,Carrasco:2021ptp}. By allowing colour factors (for example, colour traces) to non-trivially combine with kinematic factors, it was found that that all operators present in the low energy expansion of string theory can be written in colour-kinematic dual form. It would be interesting to see if the kinematic Hopf algebra is still present in this case, and to understand the modification of the evaluation maps to incorporate more general colour structures. If this is possible, it would allow re-writing the complete string EFT in terms of this novel algebra.  Similarly, one can attempt to modify the Hopf algebra even beyond adjoint type numerators, as the colour-kinematic duality is not necessarily  tied to adjoint type constructions, as recently explored in \cite{Carrasco:2022jxn}.

The BCJ numerators encoding higher dimensional corrections to Yang-Mills can be double-copied to obtain corresponding corrections to gravity \cite{Broedel:2012rc,He:2016iqi}, which could be relevant to black hole  scattering \cite{Brandhuber:2019qpg, AccettulliHuber:2020oou}. Furthermore, given that the close connection between the Hopf algebra and the geometrical perspective described in~\cite{Cao:2022vou}, it is interesting to see if a similar permutohedra construction also exists for the higher-derivative operators discussed in this paper. Finally, it was already noted \cite{Brandhuber:2021bsf, Cao:2022vou} in the case of Yang-Mills theory that BCJ numerators obtained using kinematic Hopf algebra have very similar structures of BCJ numerators from the so-called covariant double-copy approach \cite{Cheung:2021zvb}. Recently, this approach was also extended to the higher-derivative theories \cite{Bonnefoy:2023imz}; it would be very interesting explore more precise connections between these two approaches, with or without higher-derivative corrections.

\section{Acknowledgements}
We would like to thank Andreas Brandhuber and Gabriele Travaglini for stimulating discussions.  GC has received funding from the European Union's Horizon 2020 research and innovation program under the Marie Sk\l{}odowska-Curie grant agreement No.~847523 ``INTERACTIONS''.  LR is supported by the European Union Horizon 2020 Marie Sk\l{}odowska-Curie Individual Fellowship, Grant No. 101025095 ``BUFFS''. CW is supported by a Royal Society University Research Fellowship URF\textbackslash R\textbackslash 221015 and a STFC Consolidated Grant, ST\textbackslash T000686\textbackslash 1 ``Amplitudes, strings \& duality".


\appendix

\section{Six-point  pre-numerator} 
\label{sec:sixpts}
At six points, the pre-numerator is given by
\begin{align} \label{eq:six-num}
&\npre(1234,v)={1\over 4}\Big[\nn\\
&-\frac{p_1\mdot F_2\mdot v p_{1 3 2}\mdot F_4\mdot v \left(v\mdot F_{1 3}\mdot v+p_1\mdot v p_3\mdot v W(1 3)\right)}{p_1\mdot v p_{1 3}\mdot v p_{1 3 2}\mdot v}-\frac{p_1\mdot F_2\mdot v p_{1 3}\mdot F_4\mdot v \left(v\mdot F_{1 3}\mdot v+p_1\mdot v p_3\mdot v W(1 3)\right)}{p_1\mdot v p_{1 3}\mdot v p_{1 3 4}\mdot v}\nn\\
   &-\frac{p_1\mdot F_2\mdot v p_{1 2}\mdot F_3\mdot v \left(v\mdot F_{1 4}\mdot v+p_1\mdot v p_4\mdot v W(1 4)\right)}{p_1\mdot v p_{1 4}\mdot v p_{1 4 2}\mdot v}-\frac{p_1\mdot F_2\mdot v p_1\mdot F_3\mdot v \left(v\mdot F_{1 4}\mdot v+p_1\mdot v p_4\mdot v W(1 4)\right)}{p_1\mdot v p_{1 4}\mdot v p_{1 4 3}\mdot v}\nn\\
   & -\frac{p_{1 2}\mdot F_3\mdot v p_{1 2 3}\mdot F_4\mdot v \left(v\mdot F_{1 2}\mdot v+p_1\mdot v p_2\mdot v W(1 2)\right)}{p_1\mdot v p_{1 2}\mdot v p_{1 2 3}\mdot v}-\frac{p_{1 2}\mdot F_3\mdot v p_{1 2}\mdot F_4\mdot v \left(v\mdot F_{1 2}\mdot v+p_1\mdot v p_2\mdot v W(1 2)\right)}{p_1\mdot v p_{1 2}\mdot v p_{1 2 4}\mdot v}\nn\\
     & +\frac{\left(p_{1 2}\mdot F_{3 4}\mdot v+p_3\mdot p_{1 2} p_4\mdot v W(3 4)\right) \left(v\mdot F_{1 2}\mdot v+p_1\mdot v p_2\mdot v W(1 2)\right)}{p_1\mdot v p_{1 2}\mdot v}\nn\\
   &+\frac{\left(v\mdot F_{1 4}\mdot v+p_1\mdot v p_4\mdot v W(1 4)\right) \left(p_1\mdot F_{2 3}\mdot v+p_1\mdot p_2 p_3\mdot v W(2 3)\right)}{p_1\mdot v p_{1 4}\mdot v}\nn\\
   &+\frac{\left(v\mdot F_{1 3}\mdot v+p_1\mdot v p_3\mdot v W(1 3)\right) \left(p_1\mdot F_{2 4}\mdot v+p_1\mdot p_2 p_4\mdot v W(2 4)\right)}{p_1\mdot v p_{1 3}\mdot v}\nn\\
   &+\frac{p_{1 2 3}\mdot F_4\mdot v \left(v\mdot F_{1 2 3}\mdot v+p_1\mdot v p_2\mdot F_3\mdot v W(1 2)+p_3\mdot v v\mdot F_1\mdot p_2 W(2 3)+p_1\mdot v p_3\mdot v W(1 2 3)\right)}{p_1\mdot v p_{1 2 3}\mdot v}\nn\\
   &+\frac{p_{1 2}\mdot F_3\mdot v \left(v\mdot F_{1 2 4}\mdot v+p_1\mdot v p_2\mdot F_4\mdot v W(1 2)+p_4\mdot v v\mdot F_1\mdot p_2 W(2 4)+p_1\mdot v p_4\mdot v W(1 2 4)\right)}{p_1\mdot v p_{1 2 4}\mdot v}\nn\\
   &+\frac{p_1\mdot F_2\mdot v \left(v\mdot F_{1 3 4}\mdot v+p_1\mdot v p_3\mdot F_4\mdot v W(1 3)+p_4\mdot v v\mdot F_1\mdot p_3 W(3 4)+p_1\mdot v p_4\mdot v W(1 3 4)\right)}{p_1\mdot v p_{1 3 4}\mdot v}\nn\\
   &-\frac{1}{p_1\mdot v}\Big(v\mdot F_{1 2 3 4}\mdot v+p_1\mdot v p_2\mdot F_{3 4}\mdot v W(1 2)+v\mdot F_1\mdot p_2 p_3\mdot F_4\mdot v W(2 3)+p_4\mdot v v\mdot F_{1 2}\mdot p_3 W(3 4)\nn\\
   &+p_1\mdot v p_3\mdot F_4\mdot v W(1 2 3)+p_4\mdot v v\mdot F_1\mdot p_2 W(2 3 4)+p_1\mdot v p_4\mdot v W(1 2 3 4)\Big)\Big]\,. \nn\\
\end{align}
One can generate more examples of the BCJ numerators and explicit form of the $W$ functions in {\href{https://github.com/AmplitudeGravity/kinematicHopfAlgebra}{{\it kinematicHopfAlgebra} GitHub repository}}~\cite{ChenGitHub}.

\bibliographystyle{JHEP}
\bibliography{KinematicAlgebra}

\end{document}